# Developing Strategies to Increase Capacity in AI Education

Results of the LEVEL UP AI Roundtable Discussions

**September 2025**

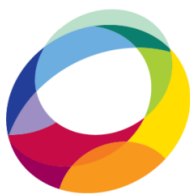

## Authors

Noah Q. Cowit, Computing Research Association

Sri Yash Tadimalla, Computing Research Association

Stephanie T. Jones, Computing Research Association

Mary Lou Maher, Computing Research Association

Tracy Camp, Computing Research Association

Enrico Pontelli, New Mexico State University

## Suggested Citation

### About the Computing Research Association (CRA)

The Computing Research Association (CRA) represents nearly 300 North American academic units, laboratories, centers, and companies engaged in computing research. Since its founding in 1972, CRA has brought together academia, industry, and government to strengthen the computing research community and its contributions to society.

CRA's mission is to catalyze computing research by leading the community, informing policymakers and the public, and promoting the development of an innovative and responsible computing research workforce. Through its programs and initiatives, CRA supports researchers across career stages and helps shape the future directions of the field.

# TABLE OF CONTENTS

# EXECUTIVE SUMMARY

Many institutions are currently grappling with teaching artificial intelligence (AI) in the face of growing demand and relevance in our world. This report discusses the first round of convening of the LEVEL UP AI project to advance AI education at undergraduate institutions throughout the United States. The Computing Research Association (CRA) has conducted 32 moderated virtual roundtable discussions of 202 experts committed to improving AI education. These discussions slot into four focus areas (1) AI Knowledge Areas and Pedagogy, (2) Infrastructure Challenges in AI Education, (3) Strategies to Increase Capacity in AI Education, and (4) AI Education for All. Roundtables were organized around institution type (e.g., R1, R2, Only Undergraduate, Community College, etc.) to consider the particular goals and resources of different AI education environments. We synthesized the thoughtful and informed roundtable discussions of experts in this report, both with respect to general AI education and to the particular needs of different institutions and positionalities. Together, these summarize the practices, challenges, and strategies institutions and individuals are exploring to improve AI education.

In particular, we identified the following high-level community needs to increase capacity in AI education. A significant digital divide creates major infrastructure hurdles, especially for smaller and under-resourced institutions. These challenges manifest as a shortage of faculty with AI expertise, who also face limited time for reskilling; a lack of computational infrastructure for students and faculty to develop and test AI models; and insufficient institutional technical support. Compounding these issues is the large burden associated with updating curricula and creating new programs. To address the faculty gap, accessible and continuous professional development is crucial for faculty to learn about AI and its ethical dimensions. This support is particularly needed for under-resourced institutions and must extend to faculty both within and outside of computing programs to ensure all students have access to AI education. Further recommendations include specific curricular and community-based solutions. To streamline learning pathways, launching an Applied Math for AI course could avoid several burdensome prerequisites, while embedding ethics throughout the AI curriculum is essential. Pedagogically, using in-class assignments can provide valuable insight into student learning processes. Creating faculty learning communities can foster a supportive AI educational ecosystem and can bolster professional development, while establishing student clubs and makerspaces can provide vital, less formal environments for AI exploration and learning.

Finally, we have compiled and organized a list of resources that our participant experts mentioned throughout this study. These resources contribute to a frequent request heard during the roundtables: a central repository of AI education resources for institutions to freely use across higher education. This initial report is intended to inform planning for additional convenings, making up the second phase of the LEVEL UP AI project, specifically to build a nation-wide AI education community and establish expert consensus around best practices for AI education. We hope that our detailed findings and resources will lead to a robust and informed discussion on how to best serve our students as we adapt to the rapidly changing space in computing education to create an AI-ready workforce.

## 1. INTRODUCTION

AI technologies are having a pervasive impact on individuals, societies, and communities. Research shows that student access to advances in AI has both positive and negative impacts on student learning [9]. Given the potential of these new technologies to transform a broad range of human activities, there is a need to increase access and capacity for AI education. Even though the U.S. is a leader in the production of AI technologies, the number of post-secondary AI courses, programs, and majors are only beginning to grow to train the needed workforce (Figures 1, 2, and 3) [11, 26]. Introducing formal educational opportunities will increase the rates of AI literacy and provide students with the skills and tools to succeed in the future AI-enhanced workforce. There are several challenges that underscore the importance of developing strategies to increase capacity in AI education, including AI workforce demands, narrow visions for AI, lack of AI teaching talent and infrastructures, and the ethical and societal implications of AI [19]. There is a growing demand for AI professionals across various sectors, including technology, healthcare, agriculture, finance, manufacturing, and government. A significant challenge is the shortage of skilled individuals with the requisite AI knowledge and expertise to perform these professional roles [28]. Increasing capacity in AI education is crucial to meet workforce demands now and in the future.

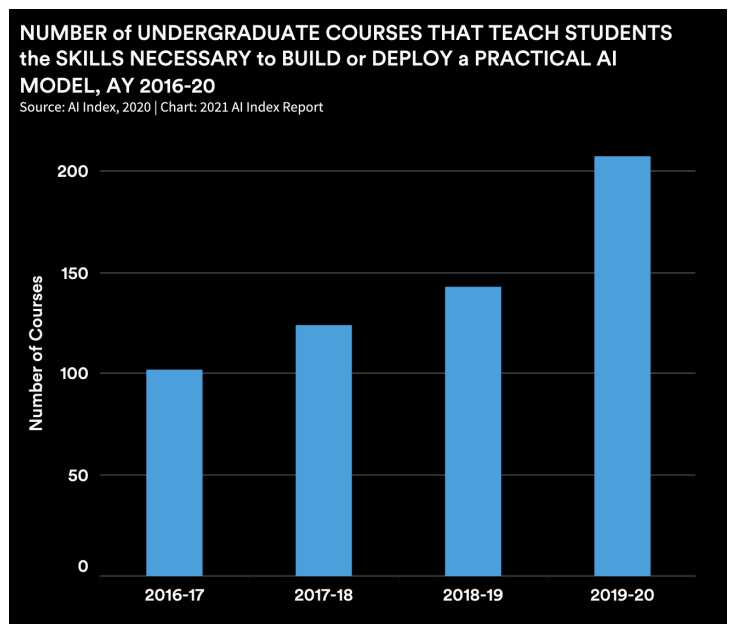

**Figure 1**: The Number of Undergraduate AI Courses Offered in the United States [26]

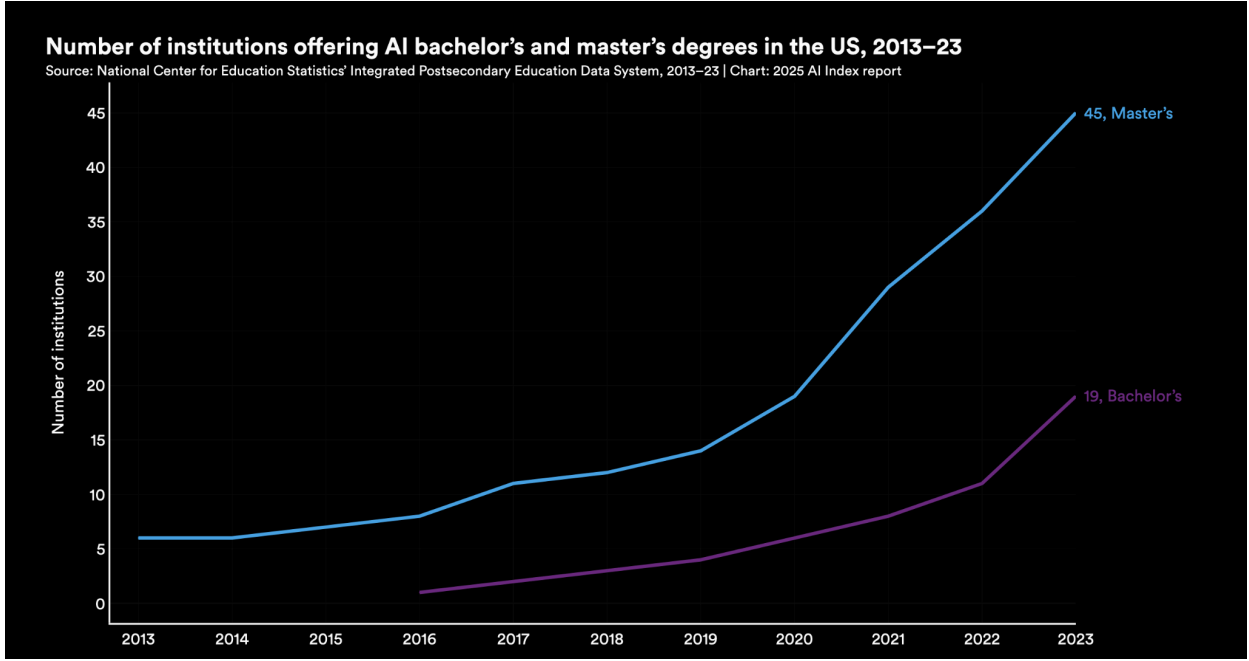

**Figure 2**: The Number of US Institutions Offering AI Study Programs [11]

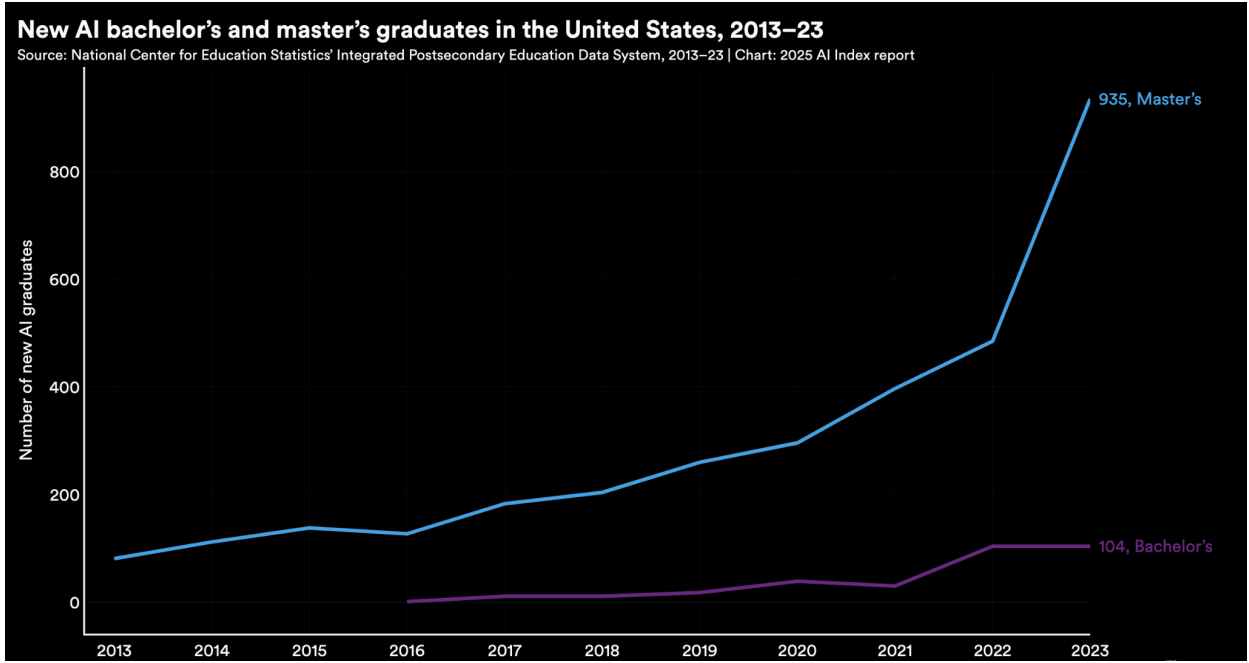

**Figure 3**: The Number of AI Bachelor's and Master's Graduates in the United States [11]

The field of AI has been criticized for how difficult it is for individuals to access and contribute to AI systems [20]. Addressing this issue requires intentional efforts to expand capacity in AI education and opportunities. Disparities in access to quality education are most prevalent in under-resourced communities [3]. Many individuals — particularly those from these underserved

communities — lack access to STEM education [16] due to factors including limited resources, inadequate infrastructure, and geographical barriers; these same factors limit access to AI education. Increasing capacity in AI education can help bridge this gap and promote social mobility [6].

Several institutions are developing new approaches to teaching AI that include ethical and societal implications [21]. Addressing these challenges requires collaborative efforts from the many stakeholders in post-secondary educational institutions, industry, and professional society organizations in the United States to develop comprehensive strategies that enhance capacity and promote access to AI education. While educational technology and generative AI can play an important role in providing access to AI education [18], this report takes steps to understand the best paths forward for comprehensive post-secondary AI learning ecosystems in the United States. That is, this report is primarily concerned with the teaching of technical AI (including AI ethics), rather than the application of AI in educational systems. As our discussions suggest, easing the capacity strains of post-secondary institutions to provide AI education and ensure AI for All may require a *paradigm shift* for how AI education is supported and conducted within educational systems [17]. This report, through its discussion of 202 experts participating in 32 roundtables on four major themes essential for improving AI education in America, provides preliminary insights for this paradigm shift.

A survey conducted by Stanford University in 2021 gathered data about the number of undergraduate courses that teach AI skills, introductory AI and Machine Learning courses, and graduate courses that focus on AI skills, as well as the number of faculty members who focus on AI research [26]. One of the highlights of the report is the large-scale increase in demand for introductory AI courses over a 5-year period, although this paralleled a similar increase in demand for computing courses writ large during the same period, making it difficult to tell if AI in particular was seeing distinctive student interest. A more recent datapoint was reported by CRA in 2024, via a preliminary survey on the *Landscape of Undergraduate Education in Artificial Intelligence (AI), Machine Learning (ML), and Robotics* [25]. To ensure a comprehensive definition of AI, the survey asked questions about courses in AI, ML, and Robotics, reported here simply as AI. The survey was distributed online to 149 academic units, all of which typically award over 200 undergraduate degrees [23] in computing each year. Of the 149 units invited to complete the survey, 56 units submitted data (38% response rate). A few highlights from the survey results are:

- **A limited number of institutions *required* an AI course in their computing degree program:** 60 percent of the survey respondents did not require an AI course for their general undergraduate computing degree program.
- **Few AI courses were offered in computing degree programs:** only 34 percent of the respondents offered more than two distinct AI undergraduate courses in their academic unit that academic year (2023-24).

*Computing Research Association (CRA)*    6

- **The need for more AI courses:** 61 percent of the survey respondents stated that more courses/sections of AI/ML/Robotics should be offered at their institution.
- **The need for more faculty that can teach AI:** 95 percent of the survey respondents identified the top obstacle to increasing capacity as the lack of faculty available to teach more courses/sections in the AI areas.

These numbers highlight the ongoing demand for increasing the pedagogical and professorial capacity to teach AI courses in computing departments. The need for increasing capacity expands by requests to integrate AI outside of computing in other disciplines. In addition to the demand for more AI courses in a computing degree, there is a need to articulate AI knowledge areas and strategies in other disciplines in order to expand the capacity for students to learn about AI and apply AI as a general problem solving instrument.

The ACM Computer Science curriculum for 2023 (CS2023) [7] includes a description of AI Knowledge Areas as well as descriptions of curricular practices in AI education. An example of an approach to increasing participation in AI is integrating AI across the institution, a strategy being developed at multiple universities, (e.g., at the University of Florida [21] and Northeastern University [2]). Considering the AI ecosystem and the impact of the lack of socioeconomic and cultural difference in the creators of AI provides a strong argument for a more comprehensive and inclusive AI curriculum [22]. Increasing participation in AI education requires a comprehensive review of goals, policies, strategies, and curriculum needs at the institutional level [10].

By considering the different needs of individuals when developing strategies to increase capacity in AI education, educators, policymakers, and stakeholders can design more responsive, fair, and effective interventions that meet the needs of all learners. The needs of students in AI education are multifaceted and dynamic, reflecting the varying backgrounds, aspirations, assets, and challenges they bring to the learning environment. This report builds on the experiences of the academic community and the report from NSF Award #2330257: Conference: NSF Workshop: Expanding Capacity and Diversity in AI Education [10] to develop a comprehensive accounting of common practices, strategies, and challenges in increasing the capacity of AI education across institutional contexts. In this project, we grouped the practices, strategies, and challenges into four major interrelated themes (1) AI Knowledge Areas and Pedagogical Practices, (2) Building out AI Infrastructure, (3) Increasing AI Educational Capacity, and (4) AI Education For All.

This early draft report is structured to guide the reader from our research methodology to our thematic findings and, ultimately, our recommendations for the future. We begin in Section 2 by outlining our methodology for data collection and analysis, which formed the basis for identifying our four central themes that organize our results. Section 3 presents these comprehensive thematic findings, with dedicated discussions on AI knowledge areas and pedagogy, infrastructure challenges, strategies for increasing educational capacity, and the imperative for "AI Education for All." Following this presentation of data, Section 4 analyzes our findings through the practical lens of different institutional types — from R1 universities to community colleges to industry partners — in order to explore tailored strategies. In Sections 4 and 5, we consider how the different institutional backgrounds and positionalities of our participants influenced responses to themes in roundtables. This report concludes in Section 6 with a summary of our key takeaways and a forward-looking set of actionable next steps for stakeholders across the AI education ecosystem.

## 2. METHODOLOGY

This report covers Phase 1 of the NSF LEVEL UP AI project that is being conducted by CRA [14]. The central data collection mechanism for this report were 32 virtual roundtables of 202 experts committed to developing AI Education from varied backgrounds. Our participants — called "experts" from this point onwards — came from varied institutions, and each provided a unique perspective on the current practices of AI educators, the challenges they currently face, and the strategies they use to overcome these challenges. The second phase of the NSF LEVEL UP AI project will provide an opportunity for more extensive community building and discussions through in-person workshops to address the goals of the project in more detail. This phased process allows consensus to build over time rather than in a single event, thereby creating a more sustainable result.

### 2.1. Data Collection

Recruitment for this event was conducted via a purposive sampling strategy to ensure a diverse and representative group of stakeholders involved in AI education. Targeted outreach was performed to solicit participation from individuals in specific roles (faculty, administrators, curriculum designers) across a spectrum of postsecondary institutions. The recruitment and registration process collected data on each participant's institutional affiliation, which were categorized as R1 (Established), R1 (Newly Classified), R2 University, undergraduate-focused college ("UGrad"), Minority-Serving Institution ("MSI"), or Community College ("CC"). The R1 category was broken up into Established and Newly Classified due to recent updates to the Carnegie Classification system, in which any institution with at least $50 million in research expenditures and awards a minimum of 70 research doctorates annually were classified as R1, leading to an additional 41 schools entering the category, a 28 percent increase [1]. These newly classified R1 institutions may have different needs than more established research institutions, which we aimed to capture in our roundtables. Additionally, key individuals from industry and the non-profit sector were invited to provide external perspectives.

The focus groups were structured as a series of concurrent roundtable discussions. Experts were pre-assigned to small focus groups based on their institutional contexts and professional backgrounds (e.g., all community college professors were placed in the same focus group). This was done so that groups of experts could speak to each other about shared challenges and opportunities at their similar institutions, as well as to avoid experts associated with institutions with high financial and research capacity (such as R1 institutions) dominating the roundtable discussion. Institution types that were represented by a small number of experts (e.g., non-profits) were only invited to participate in roundtable talks for the themes more relevant to them.

To ensure that each of the eight roundtable discussions was both productive and systematically documented, each session was managed by a two-person team: a moderator and a note-taker. These roles were distinct, with clear responsibilities designed to allow the moderator to focus entirely on facilitating the conversation, while the note-taker focused exclusively on collecting the discussion. The facilitator's primary role was to introduce a series of guiding questions (based on a script), maintain the flow of conversation, and encourage participation from all members. The moderator's script provided verbatim text for session introductions and conclusions, the precise phrasing of the guiding questions for each theme, and clear instructions for managing time and transitioning between topics. Furthermore, it included a toolkit of pre-approved, targeted probing questions designed to elicit deeper insights while reducing introduction of moderator bias. A designated note-taker was assigned to each roundtable to document the key points, emerging themes, resources mentioned, and salient quotes from the discussion. Complementing this, the note-taker's protocol standardized the data capture process through a uniform document template.

The virtual meetings were recorded and a transcript was generated from the recording for each group. This structured yet conversational process was designed to foster a rich dialogue and capture a wide range of perspectives systematically. The specific guiding questions that structured the discussions, the analytical themes that emerged, and the number of experts in each roundtable are presented in Table 1. As the expert categories can overlap — for example there are institutions which are both R1 and MSI — Table 1 represents how the experts were grouped for each roundtable session. No roundtable was made up of more than 10 participant experts; for some institution types there were several roundtables per theme (e.g., theme 1, R1 (Established)). In total, 130 post-secondary educational institutions and 12 companies/non-profits were represented in this sample, from 41 U.S. states. Roundtable attendees were selected based on their AI expertise or for their involvement in institutional AI education efforts varying from teaching an AI course, running an AI major, and/or administrating AI curriculum change.

| Roundtable Theme | Questions | Stakeholder Background | Number of Participants |
|---|---|---|---|
| **Theme 1:** AI Knowledge Areas & Pedagogy | - What AI learning content/experiences should be in the core (minimum content) of a Bachelor of AI degree in a department or college of computing?<br>- What AI learning content/experiences (i.e., any interaction, course, program, or other experience in which learning takes place) should be the core (minimum content) in a Bachelor in computing degree?<br>- What specific pedagogical approaches (i.e., methods and strategies that educators use to facilitate learning and create effective learning experiences) do you use in teaching AI? | R1(Established) | 29 |
| | | R1(Newly Classified) | 6 |
| | | R2 | 5 |
| | | Undergrad Only | 13 |
| | | MSI | 5 |
| | | Industry | 6 |
| **Theme 2:** Infrastructure Challenges in AI Education | - What resources are needed to increase the availability of AI content in undergraduate computing degrees at your institution? What are your recommendations for achieving the needed AI Education Infrastructure?<br>- What curricular resources are needed to increase the availability of AI content at your institution?<br>- What resources are needed for faculty at your institution to increase the availability of AI content (additional faculty, student assistants, professional development, release time, etc.)?<br>- What lab resources are needed at your institution to increase the availability of AI content (equipment, software, data sets, access to AI models, professional staff, etc.)? | R1(Established) | 7 |
| | | R1(Newly Classified) | 3 |
| | | R2 | 6 |
| | | Undergrad Only | 8 |
| | | Community College | 7 |
| | | MSI | 8 |
| | | Non-Profit | 6 |
| **Theme 3:** Strategies to Increase Capacity in AI Education | What are good practices for increasing capacity to AI courses in undergraduate computing programs (thinking from the student's lens)?<br>- Capacity/Access to AI courses in the computing major?<br>- Capacity/Access to AI courses in a computing minor?<br>- Capacity/Access to an AI major?<br>- Capacity/Access to AI courses for students in a non-computing major (X+AI, certificates, etc.)? | R1(Established) | 27 |
| | | R1(Newly Classified) | 6 |
| | | R2 | 4 |
| | | Undergrad Only | 5 |
| | | MSI | 6 |
| **Theme 4:** AI Education for All | What principles and resources are needed to enable/attract students from all backgrounds to be successful in AI courses and to learn to use AI responsibly? Focus on specific needs at your institution, or describe principles and resources at your institution that are a model for others.<br>- Curriculum paths and prerequisites?<br>- Introductory AI content?<br>- Pedagogical approaches?<br>- Socio-technical content?<br>- Accessibility issues? | R1(Established) | 20 |
| | | R1(Newly Classified) | 6 |
| | | R2 | 3 |
| | | Undergrad Only | 9 |
| | | MSI | 6 |
| | | Non-Profit (included in R1 roundtable) | 1 |
| Total Participants | | | 202 |

**Table 1.** Round Table themes, questions, and expert quantities by institutional grouping.

## 2.2. Data Analysis

The foundation for the analysis was a comprehensive data set collected from each of the 75-minute roundtable sessions listed in Table 1. The data set for analysis were created in three categories composed of four distinct sources: (a) the transcript generated from the video recording, capturing the full dialogue and typically ranging from 10,000 to 15,000 words; (b) the structured notes produced by the designated note-taker, which summarized key arguments and captured salient quotes, averaging between 2,000 and 4,000 words; (c) the complete text from the session's chat window, containing expert questions and shared links (appended to the notes); and (d) a curated list of resources, such as links to articles and tools, provided by the participating experts. The data from all eight sessions across the three categories was aggregated and analyzed using a two-stage, human-AI hybrid analysis process. First, the entire data set was processed using Google's Gemini 2.5 (pro), which was prompted to generate a structured summary for the guiding questions under each theme, featuring an overall summary, distinct narratives for sub-questions, and granular sub-narratives reflecting specifics for demographic groups.

For the prompt creation, in addition to using precise language and offering clear instructions to the LLM, our standardized queries for each prompt instructed Gemini to generate responses using the content in the data provided directly to it (see prompt below). This methodology — inspired by retrieval augmented generation — can reduce hallucinations by providing LLMs with a curated set of relevant and factual documents alongside the user's prompt [4]. By grounding the model in an external knowledge base, Gemini generated responses are based on the provided documents and are less likely to draw on the pre-trained, internal, and sometimes flawed, parametric language model, reducing the likelihood of fabricating information.

The prompt used:

*"You are an expert qualitative researcher. Please analyze the following transcript of a roundtable discussion. Your analysis should be structured as follows:*
*Begin with a concise summary of the main ideas, key discussion points for each of the guiding questions, and an overview of the overall conversation flow.*
*For each of the main guiding questions presented in the transcript (e.g., "Question 1:...", "Question 2:...", etc.), please perform the following analysis:*
- *Identify and list the key themes that emerged in the discussion related to that specific question.*
- *For each theme, provide a detailed synthesis of the participants' views, arguments, and contributions. Capture the consensus, disagreements, and nuances of the conversation.*
- *Support your analysis with direct, but anonymous, quotes from the transcript to ground your findings in the source material.*

*In a separate section, analyze how participants from different institutional contexts (e.g., R1 University, Small College, MSI, Industry) framed their contributions. Describe how their institutional positionality appeared to shape their perspectives, priorities, and concerns.*
*In a final, separate section, identify any insights or reflections that appear to be shaped by aspects of personal identity or professional positionality (e.g., race, gender, academic background, years of experience). Only include what is clearly stated or strongly implied by participants in the transcript. Use anonymous quotes to support your analysis.*

11   *Developing Strategies to Increase Capacity in AI Education*

> ***Ground all analysis strictly in the provided transcript.*** *Do not add, infer, or speculate on information that is not explicitly stated or clearly implied by the participants.*
> ***There is no word limit*** *— provide as much detail as necessary to capture the full meaning and nuance of the conversation. Do not access or reference any external websites or online sources. All analysis must be based solely on the transcript content and the provided resources in the notes."*

To evaluate the coverage of AI-generated summaries of roundtable transcripts, we employed a human-centered spot-check methodology grounded in qualitative validation. Full transcripts were segmented by timestamp into one-minute intervals, from which a stratified random sample of segments was selected. For each sampled segment, two human AI education expert coders reviewed the transcript text, identified key themes or concerns expressed by the speaker(s), and recorded these using an open coding approach guided generally by generated themes from the AI summaries. These identified themes were then cross-referenced against the AI-generated summary to determine whether the themes were fully present, partially addressed, or absent. The coders documented both the judgment and the corresponding location (if applicable) in the summary. To ensure reliability, at least two reviewers independently assessed each segment, and discrepancies were discussed to reach consensus. This method draws from and adapts segment-level validation approaches in prior summarization evaluation literature [5, 24], allowing for a structured and replicable approach to assess alignment between source content and AI-generated output [8, 15, 27].

In the analysis stage, the authors treated the AI generated structured summaries as a thematic output, systematically reviewing the AI-generated narratives to synthesize, validate, and refine the core findings into a set of key messages. Although we can be reasonably confident most of the AI summaries used as part of our qualitative analysis are grounded in roundtable data, we have concerns regarding hallucinations in the model. While we have made efforts to reduce the prevalence of hallucinations (noted above), we are unable to guarantee they are fully excluded. However — with hallucinatory limitations in mind — using AI in this case was selected for its utility in quickly drawing attention to the general flow of conversation, which is the current unit of analysis. A third author also manually reviewed and read all transcripts in order to offer guidance and nuance on any summaries. These recommendations are not assertions of best practices or uncontested truths. Instead, **they are intended to inspire further discussion among domain experts**.

Additionally, it is important to note that while we are inspired/informed by roundtable discussions and AI summaries of transcripts, what we present is based on our interpretations of the data collected and the AI generated summaries. All findings are filtered through our varied understandings of process, data, and topic. Section 3 presents the key messages derived from this analytical process.

# 3. THEMATIC FINDINGS

## 3.1. AI Knowledge Areas and Pedagogy

| Key Recommendations |
|---|
| 1. **AI Literacy:** Experts advocate for a foundational AI literacy that includes an understanding of ethics, core concepts, and societal impact. A dedicated AI bachelor's/major builds on AI literacy with a robust mathematical foundation, a deep dive into both symbolic and statistical AI, an in-depth knowledge of the computational aspects of AI, and ethical implications of data practices.
| 2. **Applied Experiential Learning:** A strong emphasis was placed on ensuring students can apply their AI knowledge in specific domains and that they have experience doing so through a hands-on capstone experience.
| 3. **An Everchanging Curriculum:** Experts expressed that AI (and thus AI education) is a continuously developing discipline, and that AI curriculum must be flexible to be successful.
| 4. **Scaffolding AI Knowledge:** Given the complexity of Artificial Intelligence approaches, different types of scaffolding were seen as critical pedagogical strategies. These included curriculum scaffolding, assignment scaffolding, and deconstruction.
| 5. **Use of AI in Programming Education:** A key debate emerged about the disadvantages and advantages of AI tools in introductory courses. The recommendations ranged from prohibiting to encouraging AI coding tools in introductory courses, with more consensus on teaching students how to critically analyze and debug AI-generated code in upper-level courses. |

Roundtable participants, all of whom were committed to developing AI education ( referred to in this report as "experts"), had overlapping ideas of essential AI knowledge in dedicated degree programs as compared to within a broader computing major. For instance, experts supported conversation and coursework about ethics, responsibility, and societal impact of AI in both computing and dedicated AI majors. Additionally, they advocated for an interdisciplinary approach to AI education to equip students with the skills to work with domain experts from fields like healthcare, finance, or the humanities. For any kind of AI education, experts recommended that students be given the following five learning opportunities alongside ethics:

1. A brief history of all forms of AI
2. An introduction to symbolic AI (e.g., search, logic)
3. An introduction to statistical AI (e.g., core concepts of ML, classification, regression)
4. An overview of major AI application areas (e.g., NLP, computer vision).
5. A standard computing core of programming, data structures, and algorithms.

Some experts also added computer systems and hardware to this list, arguing that understanding how to use hardware efficiently is critical for machine learning at scale.

Experts identified particular considerations when teaching AI within broader computing majors. They emphasized that every computing graduate, regardless of their specialization, must have basic AI literacy. In that, graduates are able to understand what AI is, what it is not, and how to reason about its capabilities and limitations. In particular, experts highlighted the need for a foundational AI course to become a required part of the core computing curriculum, rather than solely an upper-level elective. Many noted their departments were already moving in this direction, driven by student demand and the ubiquity of AI. They also noted a variety of approaches to this matter, not limiting students to one kind of AI like LLMs, but a variety of models suited for students interests and ethical engagements. Experts also highlighted the opportunities and challenges involved in integrating AI tools and concepts across the computer science curriculum, in particular with respect to AI-assisted programming. For many (though not all) the consensus was to prohibit AI tools in introductory programming courses (CS1/CS2) to ensure students build necessary foundational skills, but to actively teach and encourage their use in more advanced courses like software engineering. Experts generally thought that the focus could shift throughout a computer science curriculum from writing code from scratch to evaluating, debugging, and integrating AI-generated code.

The coursework recommended for a dedicated AI major diverged quite significantly from that of an AI education within computer science, in particular regarding the breadth and depth of the mathematical, computational, and ethical topics covered. For instance, many experts agreed that a dedicated Bachelor of AI degree must be built upon a robust mathematical foundation. Experts consistently identified linear algebra, calculus (often multi-variable), probability, discrete math, and statistics as highly recommended or "non-negotiable." However, there was considerable conversation about the best way for these key mathematical areas to be introduced to students. Some experts believed that these courses should be taught as traditional, rigorous math courses taken early in the curriculum. Others — particularly from institutions with a focus on applied learning — worried that this "front-loading" of challenging theoretical mathematics courses would dissuade promising students from starting or continuing the AI major. This led to the popular suggestion of creating specialized courses like "Math for AI," which would teach the necessary mathematical concepts in a more integrated and concise manner, while providing students with immediate AI application examples to sustain their engagement in the AI content.

Additionally, experts recommended an AI specific major to include a deeper dive into the basics of AI, with focus on "Good Old-Fashioned AI" (GOFAI), including symbolic reasoning, search, planning, logic, and knowledge representation (e.g., knowledge graphs). For example, one expert stated:

> "*I will fight till the end to make sure that [symbolic AI] is included. This is not going to be a degree in machine learning... if this is AI, then we have to include the broad umbrella of AI.*"

To re-iterate, a variety of component areas of AI were also mentioned as needing to be taught in considerable depth, including Machine Learning (supervised, unsupervised, and reinforcement learning), Deep Learning (foundational concepts of neural networks, and specific architectures like convolutional neural networks and transformers), and core applications of AI (e.g., natural language processing, computer vision).

Many experts agreed that any AI curriculum should be designed to weave considerations of fairness, accountability, transparency, bias, privacy, and societal impact into technical courses. The discussion went beyond a narrow definition of "bias in data sets" to include a systemic, lifecycle view of AI's impact, covering aspects like:

- Data Provenance: Understanding where data comes from, how it's labeled, and the human labor involved.
- Environmental Costs: The energy consumption and environmental footprint of training large models.
- Societal and Community Impact: The effect of data centers on local communities and the labor behind models.

It was regularly mentioned by our experts that ethics could not be an afterthought or a single, isolated course, but that it must be present at every step in a larger AI curriculum: "*I think it is fundamental to be at every level and every course thinking about ethics, not just have it all put into one class... It should be infused throughout the entire curriculum.*" A strong emphasis was placed on ensuring students can apply their AI knowledge and that they have experience doing so. This included creating tangible exercises where students could take the concepts of ethics and apply it to interpreting a particular data set or take an ethical AI framework and use it to evaluate a problem. Alumni of general computing programs discussed how the one credit or elective course they took on ethics left them underprepared for handling these matters on the job. A holistic and integrated approach to ethics and societal impacts–as a foundational sequence and incorporated into all courses of AI education programs–may be one way to rectify this.

The idea of a capstone project or a similar culminating experience was widely endorsed. These projects were described as "end-to-end," requiring students to define a problem, gather and clean data, build a solution, and evaluate its impact. One expert explained this in depth:

> "*They really need to have this hands-on experience... it'll be very important for them to have a practicum... something where they really have an end-to-end* *experience from defining the problem and gathering the information... and then having a final solution that they will have to present."*

The concept of an "AI development lifecycle" or "MLOps" (Machine Learning Operations) was also introduced, with experts arguing that students need to understand the full process from data ingestion and cleaning to model deployment, monitoring, and maintenance at scale.

For pedagogical strategies for teaching AI courses, experts frequently expressed a desire for students to have hands-on, project-based, learning experiences (e.g., implementing an algorithm from scratch, developing end-to-end applications). Building tangible systems was highlighted as a way of helping students learn most effectively while "*kind of tasting a little bit of what AI can actually do.*" Gamification was considered as a pedagogical strategy. Some ideas floated included creating class-wide tournaments for game-playing AI agents or using leaderboards for machine learning competitions. Another key element was the use of team-based projects, which help develop crucial "soft" skills like communication, collaboration, and project management.

Case studies were also recommended as a potential pedagogical strategy, particularly to teach the key ideas of ethics and societal impact. Experts recommended using examples that are local or personally relevant to students to increase engagement. Using ill-defined problems was another strategy that invites students to engage in critical thinking, creativity, and demonstrate their problem-solving process rather than following a prescribed recipe. While experts were eager to include ethics in their AI curricula, some expressed that they and their colleagues had a lack of training in how to teach ethics successfully. Faculty suggested course-releases as an opportunity for training around updating their AI curriculum and how to effectively teach ethics throughout. Further, they suggested that time dedicated to interdisciplinary collaboration with faculty in other departments would be beneficial in creating joint curriculum and approaches, particularly looking towards humanities and social sciences scholars in Philosophy, Sociology, and Theology.

Given the complexity of Artificial Intelligence approaches, different types of scaffolding were seen as critical pedagogical strategies. These included curriculum scaffolding (introducing a concept at a high level early on and then revisiting it with more depth in later courses), assignment scaffolding (providing skeleton code or Parson's problems to decrease student cognitive load), and deconstruction (starting with a complex, real-world application such as AlphaGo, and working backward to deconstruct the components).

Finally, experts discussed the use of generative AI in the process of teaching AI. Experts spoke about how banning AI tools entirely can be impractical and counterproductive to students. They suggested several strategies which could allow these tools to complement rather than inhibit student learning. For instance, experts suggested "show your work" policies relating to the use of AI on projects, asking students to document their thought processes when creating code, including their use of AI prompts. Experts also suggested a shift from laboratory to in-class/oral assessment methods, such that students must demonstrate they understand key material without the assistance of AI. For more advanced students, experts recommended shifting the focus from writing to evaluating code, allowing students to create draft code with AI but requiring them to critique, debug, test, and analyze it for accuracy, practicality, and efficiency:

> *"I tell them, no, use ChatGPT. And now you have to criticize that and you have to analyze it and go back and change your prompt and get something else. You have to figure out what is wrong and what is right."*

## 3.2 Infrastructure Challenges in AI Education

| Key Recommendations |
|---|
| 1. **A Broad Definition for AI Education Infrastructure:** Delivering quality higher education on AI is hampered by infrastructure challenges that extend beyond hardware to include faculty talent, support for professional development, curriculum, IT staff, and data resources.<br>2. **Collaborating Across the Institutional Divide:** Experts advocated for breaking down institutional silos in order to alleviate some of the challenges involved in developing AI curricula, teaching resources, and access to data and models.<br>3. **Competitive Resource Landscape:** Faculty and staff are difficult to hire due to intense competition between academic institutions and from industry, and the varying resources different employers can offer to prospective talent. Experts expressed concern about uncompensated and unrewarded curricular development work, deprioritizing educational innovation.<br>4. **In-house Computing or Cloud Services?** A critical shortage of GPU access for students limits hands-on learning, forcing a debate between costly cloud services and difficult-to-maintain local hardware with minimal dedicated-to-AI staff IT resources.<br>5. **Where is the Data?** High-quality, open-source data sets to build course materials around are hard to come by, particularly those that are related to topics meaningful to students. |

A common opinion among our experts was that AI (and thus AI education) is a continuously developing discipline, and as such, AI infrastructure must be flexible to be successful. Experts expressed frustration at the lack of time to update curricular resources at their institutions, with one instructor stating,

> *"I teach AI... and I have to update the course every year."*

One expert described their experience designing an AI curriculum, mentioning its "*very hard to find a textbook*" for newer courses like deep learning, especially one that includes "questions and exercises." Many experts told us that they rely on "blog entries" or the goodwill of colleagues at other institutions to develop courses. The need for curricular guidelines that are updated more frequently than the traditional "decade basis" was also raised, with one expert expressing that:

> "*[AI] changes so fast that looking for a curriculum and finding one means that it's probably out of date already.*"

Faculty described "*building the plane as they're learning how to fly it,*" often without any sort of instruction or guidance.

A key recommendation mentioned in roundtables was the necessity of professional development opportunities for AI educators, to assist them in this ever-changing space. Experts highlighted that professional training must be invested into and offered by institutions which aim to teach current and high quality AI courses, and that this training be accessible and continuous. One suggestion was to create "faculty learning communities," where instructors can co-learn how to incorporate AI into their curricula and teaching. It was impressed that professional training should be contextual and discipline specific — such as helping a chemical engineering professor understand how AI is used in that field — rather than just a generic overview.

The importance of professional training and retraining was described as particularly essential due to the shortage of domain experts in AI, particularly for less competitive and smaller institutions outside of major technology hubs. Market competition for personnel resources was consistently described as very intense, and even when hiring calls were successful, experts described difficulties in retaining AI faculty hires, who often left for different — more lucrative — opportunities elsewhere. These faculty shortages and turnover rates place a heavier burden on existing faculty to retool and cover an ever-expanding field, which requires both motivation and opportunities for faculty professional development.

At many institutions, experts described the recurring barrier to this professional development occurring as a lack of compensated time. For faculty members already operating with full schedules, the expectation that they "*spend their weekends [or] spend their summers doing that thing [keeping up with the field and constantly develop new materials]*" is overwhelming, particularly as institutions often do not offer resources to compensate faculty for this additional work. It was impressed that meaningfully restructuring curricula would prove impossible without institutional investment.

In addition to a lack of monetary support from institutions for professional training and curriculum design, experts suggested that "traditional faculty metrics" were partially to blame for the lack of progress in their institutions towards cutting-edge AI education. Faculty evaluation is traditionally based on grants and publication, rather than on educational innovation. In order for curriculum development and faculty retraining to be successful, it *"has to be built into the recognition that we give faculty member[s]...promotion, tenure, annual evaluations."*

Experts advocated for breaking down institutional silos in order to alleviate some of the challenges involved in developing AI curricula and teaching resources. One expert argued for moving away from "competition to more of…collaboration" both "within an institution" and "across institutions." Internally, this could involve creating partnerships where — for example — the computer science department can work with the humanities or a center for teaching and learning to champion intentional approaches to AI education campus-wide. Externally, this may involve the creation of consortia or multi-institution networks to share resources, curriculum and expertise. Experts also expressed a desire for more "*partnerships between industry and faculty*" to stay current with workplace demands and newest advancements in the field.

While much roundtable conversation about AI infrastructure centered around the need for skilled educators and updated curricula, deficiencies in computational resources were also spoken about extensively. The need for GPU access was one concern. As one expert stated,

> "*Students want to do interesting things... And we just don't have the GPU access."*

Even institutions with high-performance computing (HPC) centers find that these resources are "*primarily being used by researchers,*" leaving educational users to "*compete for resources.*" This scarcity makes it difficult for many institutions to teach modern AI, as running and training models is computationally intensive. Many educators have turned to services like "Google Collab" as a workaround, but there was acknowledgement by experts that it "*cannot carry the weight of what we're talking about*" for more advanced projects.

There was debate among our experts as the best way to provide computational resources to students. Often noted were the resource differentials that institutions had to support compute power, support advanced AI model options, write grants, and access industry / research lab partners. Some institutions are investing in local hardware, such as a "small supercomputer" or a dedicated server to run models locally. The motivation for local hardware is often ease and control, as it gives students the ability to "*experiment and play around a little bit in a low stakes way,"* and to handle projects with "confidential information" that cannot be sent to an external API. Conversely, other institutions have "*gone completely to Google Cloud*" or AWS, arguing that managing local hardware is a "solved problem" by cloud providers and that students need "*cloud skills on an actual cloud provider*" as that is what they will use professionally. The significant downside to the cloud is cost and unpredictability. Students get "*worried that they're going to spend like $10,000 accidentally,"* and instructors have seen projects that "*blew through all the [class'] tokens in one weekend.*"

A key concern by many experts was the need for advanced non-faculty computational staff (e.g., technical support staff) to maintain the backend as well as individual computer troubleshooting. As one expert explained, GPU clusters "*by themselves are not enough unless there is somebody that knows how to use them properly.*" Institutions need "*professional staff that actually helps faculty members*" and can provide technical support and training. This is a significant challenge, as skilled IT staff are often "*poached by industry,*" leaving university departments reliant on "one person" or a faculty member serving as a non-compensated liaison as "*part of his service.*"

High-quality data sets and open-source tools were also described as an institutional need to deliver a high-quality AI education. For instance, one expert called for "*an inventory of free open source well-documented tools that can be then modified and customized,*" while another called for "*open data sets that we can use*" and "*data sets that are ready to use for the classroom.*" The ideal is not just raw data and tools, but data and tools with "*curricular infrastructure built around them, whether it be even just like a lesson plan or... Jupyter notebooks.*" There was a strong call for more inclusive data, such as "*neurodiverse data sets, linguistically diverse data sets*" so that the models students build and analyze are less likely to perpetuate bias. The challenge is that much of the best data is proprietary, as one person noted, "*industry is at the cutting edge because they have the data.*" The idea of a central repository, similar to what exists for cybersecurity, was floated as a potential solution to prevent faculty from "*reinventing what... others have already solved.*"

## 3.3. Strategies to Increase Capacity in AI Education

| Key Recommendations |
|---|
| 1. **Major Structural Changes**: Course capacity and faculty advising expertise are an acute problem across institutions. To meet overwhelming student demand, institutions are exploring strategic curricular and structural changes to increase capacity for AI education for both majors and non-majors. One recommendation is integrating AI concepts much earlier in the undergraduate computer science curriculum.
| 2. **Degree vs Concentration:** A central debate is whether to create distinct AI majors — driven by market demand — or to maintain AI as a stronger concentration within a traditional computer science degree.
| 3. **Math for AI:** The critical debate on Math for AI is about how much math should be required and how to teach the required math. There is a growing push for integrated "Math for AI" courses that prioritize application over pure theory.
| 4. **Interdisciplinary AI Courses:** An opinion raised throughout roundtables is the necessity of non-computing majors having the opportunity to learn about AI. The difficulty non-CS students face in accessing AI courses in CS departments has led to a trend of external AI |

> courses (AI+X). The balance between enrollment accessibility and instruction quality for these non-CS courses was a key discussion among experts.
> 5. **AI Outside of the Classroom:** Experts also recognize the value of informal learning ecosystems — such as student-led clubs and AI makerspaces — to broaden access and spark interest outside of formal courses.

Our experts committed to improving AI education described course capacity as an acute problem across institutions. Classes in AI were described as overbooked to a degree that "*it's just a crapshoot whether you can get in,*" even if you are a declared computer science major. For computer science minors and non-computer science students, the likelihood of gaining admittance into an AI course was described by many experts as difficult to almost impossible.

The difficulty non-CS students face in accessing AI courses has led to a notable trend: "*proliferation of these very computational AI ML courses out in other disciplines.*" One expert noted that at their university, about "*50% of them [AI courses] were completely outside of*" CS or ECE, appearing in fields like environmental engineering, social sciences, and economics. Another dean stated their university's strategy is that "*all the disciplines need to teach their own AI.*" This decentralization was seen as a double-edged sword. On one hand, experts described this as representing a healthy, distributed interest, allowing for domain-specific applications of AI. Affirmed by an opinion with much consensus throughout roundtables was the necessity of non-computing majors having the opportunity to learn about AI. Offerings of standalone AI courses or AI micro-credentials *"AI+X (economics, biology, etc.)"* were suggested as strategies of broadening the capacity of AI to students outside of the major. On the other hand, however, experts raised concerns about the quality, consistency, and well-roundedness of the AI instruction outside of CS. This has prompted some institutions to explore more formal interdisciplinary structures, like an online CS minor offered collaboratively across colleges. Co-teaching was also explored as a potential solution, where a CS faculty member partners with a faculty member from another discipline to ensure both technical accuracy and domain relevance.

Collaboration — while a potential solution — was described by some experts as a challenge in itself. Experts mentioned "academic territoriality" and hesitant departments as significant hurdles. Some institutions have tried to foster collaboration by creating university-wide structures like a "Data Science and AI Academy" or a "Data Science and AI collaborative" to serve as a neutral meeting ground. The most successful approach seems to be one of partnership rather than ownership. One expert re-framed the idea of "relinquishing" control towards "*its collaboration, its support, its extending that education and making it accessible."*

One challenge raised in teaching AI to non-computing majors was the lack of student access to advisors with domain expertise and in their professional networks. When speaking about computer science minors, one expert stated *"they are outside the primary departmental* *community and miss out on the informal peer networks that majors rely on."* One expert who advises a CS-minor explained:

> *"Students in a major benefit from the network they have in their major to hear [advice] from students... The students who come into the minor, they're clueless [to these insider tips]."*

However, experts also doubted that this advising problem could be solved if CS-minors just had a computer science advisor. One expert noted the challenges of this approach, stating:

> *"If they're in a totally different field... I might be giving them advice that's not going to help them on their path."*

One potential solution to this is better inter-departmental coordination in advising, with students potentially co-advised by faculty in multiple departments.

Like in AI Infrastructure roundtables (theme 2) and in previous publications, the lack of qualified AI faculty was identified as a major barrier to increasing capacity to meet soaring student demand. One expert detailed their institution's strategic hiring of 11 new faculty, including several in AI. Another noted, *"we convinced the dean and the provost to give us more positions to have strategic hiring…faculty in machine learning and AI."* For smaller institutions, the challenge is even greater. An expert from a department with only three faculty members stated:

> *"We face a capacity challenge. And this limited staffing makes it difficult to expand these AI offerings."*

Like theme 2 experts, computational capacity — particularly GPUs for training large models — were identified as a need to increase capacity in AI. This computational resource scarcity was described as directly impacting the ability of institutions to provide hands-on experience, particularly for undergraduate students who have difficulty accessing national resources like the NSF ACCESS program.

Curricular and pedagogical re-design was described as another avenue of increasing capacity in AI. In many institutions, AI courses remain upper-division electives rather than required courses for the computer science major. However, there is a growing recognition that this model may be insufficient. As one expert argued, waiting until junior or senior year for exposure to AI is *"too late because AI motivates them."* This led to a strong theme of integrating AI concepts much earlier in the curriculum. One popular strategy is to embed AI-related assignments into foundational courses like CS1 and CS2. Another strategy was the creation of a freshman-level "AI literacy" course, even for CS majors, to provide a foundational understanding and help students *"decide whether they want to do AI at the third year level or not."*

In designing these new introductory AI courses, experts debated the merits of different pedagogical models. The traditional "bottom-up" approach, where students learn theory and fundamentals before tackling applications, was contrasted with a "top-down" model. An expert described their planned AI major as intentionally using a top-down approach for freshmen and sophomores, starting with applications to build motivation and context. This is then complemented by the traditional bottom-up approach in the junior and senior years, where students "*understand what you heard about in the freshman year a little bit better.*" This blended model was presented as a novel and effective way to structure an entire major to maintain student engagement while ensuring foundational rigor.

One area in which there was quite a bit of disagreement among experts was whether a separate AI major — distinct from computer science — was even desirable. Some were strongly opposed, expressing concerns that it would be a "*jack of all trades... master of none*" and that employers prefer graduates with strong, fundamental computer science knowledge. One expert argued:

> "*I honestly don't see what an AI... specialist? What is that?... I don't see how you can have an effective AI major without the courses that we require for our computer science major.*"

This perspective suggests that an AI concentration within a CS degree is a more robust model. Another concern was that an AI major might be a "novelty" or "hype" that could quickly become outdated:

> "*It's going to be something else in a decade and you graduated with an AI degree that was maybe focused on LLMs. What are you going to do?.*"

Conversely, other experts saw the AI major as a necessary evolution. One argued for its importance on a national scale, stating that producing thousands of students specializing in AI is critical for "future competitiveness" in the "global competition." Others are moving forward with AI majors as a direct response to market forces, which one dean described vividly:

> "*We resisted the development of AI programs up until approximately October of last year. And that's when we realized for the first time that we have [a] hundred fewer CS freshmen enrolled. And that was enough of a signal.*"

These conversations indicate that for some colleges, the AI major is a strategic imperative for survival and relevance. This entire discussion was underpinned by an awareness of intense market pressure. Students are flocking to AI because they see high-paying jobs. As one expert stated:

> *"When these undergraduate students are coming... They always have this question of what is the return of investment."*

This demand from students and industry is a primary driver for creating AI majors. However, it also creates the "hype" that worried some experts, leading to a concern that some programs might be created "*more about the marketing*" because "*AI just sounds more exciting than the computer science.*" This highlights the challenge of balancing genuine pedagogical innovation with the powerful branding of "AI."

As in the AI Knowledge Areas (theme 1) roundtables, mathematics prerequisites were identified as a bottleneck preventing capacity in AI education from being increased. One expert noted:

> *"When you have linear algebra [and] statistics...that adds quite a bit of requirements before they [students] can actually taste what AI and ML courses are like."*

This delay was seen as making it more difficult to retain student interest in AI. Like some experts in theme 1 (section 3.1), experts in these roundtable discussions advocated for the creation of a new, integrated math course tailored for computing students (Math for AI). One university described going so far as to develop a single course in this vein to provide a more direct and applied pathway to the required mathematics skillset. The philosophy behind this course was to make it "*a math class for computer science students, rather than math class that computer science students happen to take.*" This involved focusing on intuition and application over formal proofs, using programming exercises to illustrate concepts:

> *"We're not as interested in hey, can you derive the proofs of these things? But instead, here's [a] quick python package that you can call that will compute the eigenvectors."*

Another expert echoed the need for this pedagogical shift, observing that math courses often teach students to "*grind it out with a pencil*" and fail to connect concepts to computation, lamenting that at their institution, "*there's no computer in linear algebra.*" This approach of creating bespoke, applied math courses was seen as a promising practice for increasing student retention and access in AI.

Finally, experts highlighted the immense value of learning opportunities outside of formal courses. One expert argued that "student clubs" are an overlooked but highly successful way to engage students, as they are "*their own community. It's their own leaders.*" Another championed the idea of an "AI makerspace" to allow students from all disciplines to "*play around and then go to the courses if you need...the fundamentals.*" These informal, low-stakes environments were seen as powerful tools for sparking interest, building confidence, and increasing access to the broadest possible student audience.

## 3.4. AI Education For All

| Key Recommendations |
|---|
| 1. **Strategies to Achieve AI for All:** Making AI education accessible to all requires AI for All development that focuses on building confidence and demonstrating relevance to all students.
2. **Hands on Learning:** AI for All should include hands-on learning that connects AI to authentic problems students care about, from local community connections to ethical and societal issues. AI for all is AI for good.
3. **Learning Design:** Course design should involve creating highly structured courses with frequent, low-stakes assignments and fostering a collaborative, supportive classroom climate. Several experts discussed shifting the focus of assignments from the final product to the learning process itself, incorporating metacognitive reflection.
4. **Accessibility:** True accessibility also means addressing the digital divide, proactively designing for all students — including students with disabilities — and providing faculty with training to navigate complex socio-technical conversations. |

Experts had a variety of ideas on how to enable AI for all, such that AI could be made enticing and accessible for students from a variety of backgrounds. Many of the conversations from these roundtable discussions mirrored those from the previous three themes. For instance, experts commonly expressed that the traditional math sequence was not well suited for recruiting students not as interested in mathematical theory, advocating for in-house "Math for AI" courses which are more narrowly focused to teaching students the skills they need for AI. There was also widespread agreement on the need to introduce AI concepts early in a student's undergraduate career, often through general education courses with no prerequisites.

Similar to roundtable discussions in theme 1, experts recommended relying on real-world examples whenever possible, and deconstructing complex ideas into simple, understandable components. A business school professor described their method:

> *"I always start with the single neuron simulator. I've built a little simulator that I called a naked neuron. You could see everything... in Excel... It starts from not knowing anything... to making 100% correct predictions within two-three minutes right in front of your eyes."*

This strategy uses familiar tools (Excel) and commonly required math, like algebra, to build confidence. Another expert noted the importance of project-based learning, stating:

> *"I've noticed that kind of light bulbs click... on around topics in AI literacy when we get the students hands on with building... generative AI environments."*

The goal is to show that AI is not "magic," but a set of understandable processes.

Additionally, the recurring suggestions of faculty professional development and teaching about technology ethics/societal impact were frequently raised in these roundtables. Experts made it clear that in order to be accessible, AI faculty must be familiar with the (1) societal impacts, (2) ethical considerations, and (3) technical elements of AI. As few faculty have expertise in all these areas, this likely requires some form of explicit professional training, relating to earlier conversations about educational capacity. One expert noted that among those willing to teach technical AI material, many are not trained to facilitate difficult conversations about socio-technical issues. An R1 professor argued for:

> "*structured faculty development for faculty that are teaching those courses. When you get into the socio-technical issues, a person that's only concerned about math, algorithms and programming, this is going to be difficult for them.*"

An internal ethical concern for faculty at many institutions was academic integrity. An expert from an R1 university highlighted:

> "*If faculty across the campuses don't understand the tools, then I think they can't create assessments and lectures and curriculum and content that is AI proof... or that works with AI.*"

In these cases faculty are looking for measures to identify or define what the bounds of student work looks like in the current moment. Experts also worried about how current evaluative tools may incorrectly flag students who speak English as a second language and/or are first-generation immigrants as having "cheated" with AI tools. These conversations displayed the complexities of embedded disciplinary patterns that may manifest while using "AI checking" tools given the changing educational landscape.

Experts suggested that instructors think proactively about their pedagogical practices. Some of our experts had preferences for active and experiential learning. One described their approach as "*immersive, hands-on, and body on,*" using simulations and even physical games to help students "*explore with your full bodies.*" This contrasts sharply with a passive, lecture-based model. Other experts highlighted the role course design and student evaluation can play in recruiting and retaining students. An expert from an R1 university explained:

> "*The structure of the course is pretty important, because, if it's a low structure course that only has two exams. If a student fails that midterm exam, they're going to drop the course.*"

They advocated for "*multiple ways for students to demonstrate their mastery.*" Another expert described the value of "*small weekly programming assignments*" because this "*forces the students to apply their knowledge at more regular intervals*" and makes it "*easier for the instructors to catch*" when a student is struggling. Several experts discussed shifting the focus of assignments from the final product to the learning process itself, often incorporating metacognitive reflection. For one expert, this involved asking students to document their steps and reflect on their learning:

> "*When students turn in an assignment, I ask them to... submit a reflection statement on how they use the tool, what sort of learning it enabled, what sort of thinking it maybe replaced.*"

This pedagogical approach was intended to encourage students to be more aware of their own learning and to use AI tools as a complement to their thinking, rather than a replacement for it.

Other experts highlighted the importance of a collaborative — rather than defensive — classroom climate, with the goal to build a community where students feel supported by both the instructor and their peers. Expert-recommended strategies of facilitating a collaborative classroom climate included group projects and peer learning environments. Experts made clear that these methods not only could foster a sense of community and collaboration in the classroom, but that these environments would be similar to the workplaces in which they would have to perform after graduation:

> "*You know, the fact is, once they graduate, everything they do is going to be a group project.*"

To attract a diverse group of students, experts argued that introductory content must connect to problems and domains that students care about. One expert from a small college explained their use of "*real world business projects with local businesses and nonprofits to show kind of immediate relevance of application of AI.*" An expert from an R1 university described how in the midst of the COVID-19 pandemic, "*Every single student was interested in the data that was being produced,*" and this real-world event became a powerful anchor for teaching data analysis. The key is to allow students to see the utility of AI in their own fields of interest, whether it's business, healthcare, entertainment, or societal impact. As one person put it, it's vital "*to address authentic problems that are really relevant to them [and] their lives.*"

A major concern in creating broader access to AI education was the digital divide, both in terms of students' personal computing devices, broadband internet infrastructure, access to high-end computational resources, and paid AI tools. Faculty suggested requiring students to use free or university-provided (institutional site licensed) cloud-based services to even the playing field, but also highlighted how this can create divides between institutions with more and less resources. Experts also discussed the challenges of making AI education accessible to disabled students. While experts acknowledged that AI as a whole could make learning more accessible to some disabled students (e.g., those with visual impairments) they also noted that with most emergent technologies, a significant lag exists when it comes to making those technologies accessible to disabled individuals. One expert with domain knowledge in the area stated:

> "*I don't know how well the current kind of APIs are for people who are using screen readers... usually things do not work well.*"

Another expert broadened the definition of disability, reminding the group to not only consider sensory impairments but also physical ones (e.g., "*people who can't type and... people who use not just screen readers, but also voice recognition software*"). The principle of Universal Design for Learning (UDL) was raised as a framework for creating more accessible courses from the outset.

## 4. CONSIDERATIONS FOR DIFFERENT TYPES OF INSTITUTIONS

| Key Focus Areas |
| --- |
| 1. **Research Priority:** Experts from large R1 universities detailed focuses on scale, theoretical depth, and creating new research-oriented programs, often dealing with thousands of students and internal departmental politics.<br>2. **Pedagogical Innovation:** In contrast, undergraduate and liberal arts college roundtable participants leveraged their smaller size to offer intimate, high faculty-student engagement pedagogical experiences that integrate and emphasize ethics and humanities, though they are constrained by faculty size and budgets.<br>3. **Access Oriented:** Minority Serving Institutions centered their efforts on inclusive pedagogy, addressing structural barriers like the digital divide and connecting AI to community-relevant social impact projects alongside developing workforce competitive young professionals.<br>4. **Workforce Development:** Similarly focused on practical outcomes, community college experts prioritized workforce development and industry partnerships to create direct pathways to employment.<br>5. **Adaptable AI Skills:** Industry partners stressed the need for adaptable graduates with skills in strong critical thinking and problem-solving, as well as understanding the entire socio-technical product lifecycle. Non-profit participants highlighted working to bridge gaps and scale solutions across all of these diverse institutional settings. |

### 4.1. R1 Universities

Faculty experts from research-intensive universities (R1) often framed the discussion around preparing students for graduate studies and research careers, alongside industry. They emphasized the need for deep theoretical and mathematical foundations to enable students to create new AI, not just apply existing tools. Their discussions reflected the realities of large-scale institutions, including the challenges of curriculum reform, the high demand for AI courses that outstrips teaching capacity, and the political maneuvering required to establish new degree programs. They also viewed AI as a fundamentally interdisciplinary area, frequently mentioning collaborations with other departments (engineering, philosophy, medicine) and the formation of university-wide AI task forces. Their approach to curriculum often involved creating specialized tracks or concentrations within existing CS degrees to provide depth. These institutions are often large enough to have internal battles over the "strange layout of AI machine learning courses" taught across different departments and the political complexities of establishing a new "AI and society" department and figuring out how it relates to the existing computer science department. While they may have access to more resources, like high-performance computing centers, these are often prioritized for research, forcing education to "compete for resources."

Experts from these institutions frequently framed the discussion in terms of scale and innovation. They are dealing with extensive student populations, with one expert mentioning a data science course with 1,500 students and a machine learning course with 1,800. This reality drives their solutions towards models that can operate at scale. A recurring idea was leveraging technology to overcome physical capacity constraints. One expert, an "Executive Director of Online Education," described their mindset as using online platforms to "*break all the barriers at once,*" proposing an online CS minor that could be offered not just to their own students but to a consortium of universities. The use of hybrid approaches, large lecture halls supplemented by many TAs, and open-source educational resources were all discussed as ways to manage thousands of students.

These institutions often saw themselves as needing to be at the forefront of defining new programs. The detailed proposals for standalone AI majors, complete with innovative top-down/bottom-up pedagogies and complex elective clusters, came primarily from experts at R1 schools. They also have the capacity to create entire new administrative structures, such as a "*new [redacted] College of AI, Cybersecurity and Computing*" or a "*Department of AI and Society,*" to house these initiatives.

## 4.2. Undergraduate and Liberal Arts College

Experts from 4-year institutions placed a strong emphasis on pedagogical effectiveness and high-touch student engagement. Their primary constraints were often faculty expertise and resources. One expert noted:

> "*We're a rural school... Very difficult because we usually don't have the budget for... access to many of the tools.*"

A key challenge is the immense burden on a small number of faculty who "*have to update the [AI] course every year.*" The conversation from this group often centered on the dedication of faculty who do this work out of passion, conducting "summer research" where they "*barely get paid.*" With smaller departments, offering a full, separate Bachelor of AI was seen as largely unfeasible.

> "*I think a lot of departments right now would struggle with the faculty to offer that [A Bachelor of AI] I think there are only four faculty that, maybe five, that offer courses that would sort of fall under the broad AI umbrella.*"

These institutions recognize they "*can't do it all.*" Therefore, a key strategy offered was to be clear about their educational goals and "*finding what is the niche that you think you can really do well … and focusing on that.*" This might mean focusing on a strong, applied AI course for their engineering students rather than trying to build a comprehensive AI major.

Experts at smaller institutions saw the liberal arts environment as an ideal place to integrate ethics, humanities, and social sciences into AI education, encouraged through team-teaching with faculty from other departments. There was more discussion of creative and adaptable pedagogical approaches like role-playing, debates, and using AI tools to foster peer review and build community, addressing the specific social-emotional needs of their student population. Experts often emphasized the quality of the student experience, and the ability of small institutions to offer intimate, project-based learning. An expert from a small college described their approach: "*I've been doing a lot of work on trying to use real world business projects with local businesses and nonprofits.*" The conversation was less about managing 1,000-person classes and more about creating "*meaningful connections with teaching and learning.*"

## 4.3. Minority Serving Institutions (MSI)

Experts from MSIs highlighted the importance of fairness in pedagogy. Their contributions centered on ensuring that AI education is accessible and welcoming to students from all disciplinary backgrounds. This involved being mindful of the assumptions made in course content and delivery. For example, one expert highlighted:

> "*One thing in my department that... I've had some involvement is thinking about how we teach our courses, particularly CS courses, but AI also, of how to make sure we're inclusive broadly to students of all backgrounds... that comes down to the way we're teaching our courses and how we're thinking about delivering our material.*"

The conversation in the MSI-focused groups repeatedly returned to fundamental structural issues. Experts argued that discussions about curriculum are secondary to addressing the fact that many students entering college "*don't have access to computers or internet or the digital infrastructure of the [K-12] school is not strong.*" Funding was framed not just as a resource issue but as a fairness issue. At roundtables for predominantly undergraduate universities and minority serving institutions, experts highlighted the differences in amounts of grant funding and grant writing time. There was an expressed wish for collaboration with R1 and R2 universities, who have higher funded award records, to create a net of tertiary education resources that can support multiple institutions. Further, given the changing funding landscape, there was an expression of concern that, what limited options are currently available, may disappear.

Like in other institutions, there was a strong emphasis on the need to invest in faculty at MSIs, who often have high teaching loads and serve as critical "*wraparound supports for students.*" For students, the focus was on building confidence and creating welcoming pathways, such as summer camps and near-peer mentorship programs. There was a strong emphasis on community relevance, social impact, and lowering barriers for students who may be first-generation or come from under-resourced K-12 backgrounds. For example, a professor from Hawaii directly connected their work to the local community. Another professor framed the discussion around concerns of how resources can shape "*who the winners and losers are,*" reflecting a pattern in economic and infrastructural inequalities. Experts emphasized practical, hands-on approaches like service-learning programs, where students partner with community organizations to build AI solutions. This approach not only provides valuable experience but also grounds their learning in addressing real-world community needs.

## 4.4. Community Colleges

Experts from community colleges consistently framed their contributions around workforce development, industry partnerships, and student access. This perspective centered on building confidence and demonstrating immediate relevance to a student's future career. Their primary concern was preparing students for "*potential employees for our businesses*" highlighting how many of their students expect practical courses that can aid learners professionally as they strive for economic mobility. This focus drives a need for clarity on "*what industry needs,*" leading to initiatives like creating industry focused advisory committees to understand required competencies. They grapple with whether to offer a specific "degree in AI" or to "embed it in everything," a decision tied to articulation agreements and transfer pathways to four-year institutions. Funding was a critical and recurring issue, with a focus on finding "a self-sustaining model" rather than relying on temporary grants, given that their students "certainly can't afford" to pay for expensive tools themselves.

## 4.5. Industry

Industry representatives provided a crucial external perspective, focusing on the skills and competencies they seek in new hires. Their viewpoint was shaped by the rapid pace of technological change and the practical realities of deploying AI systems at scale. These experts repeatedly stressed that technical skill alone is insufficient. Graduates must understand the entire lifecycle of an AI product, including its societal impact, ethical implications, and the importance of responsible AI principles. One industry leader said,

> "*If you have a student that doesn't care about the socio-technical aspects of the software they build, then I wouldn't hire that person... Almost everything that we do... involves teams of people … we have to think about international audiences... different cultures, different societies.*"

Given that specific tools and models become obsolete quickly, these experts valued graduates who were adaptable and had strong foundational problem-solving skills. The most critical skill to these experts in regards to generative AI tools was the ability to evaluate, validate, and debug the outputs. These leaders emphasized that they still expect graduates to engage in critical thinking and be able to make informed choices when it comes to how their technical products will affect the business.

## 4.6. Non-Profits and Ed-Tech Partners

Experts from outside traditional academic departments brought a perspective focused on bridging gaps, scaling solutions, and understanding ecosystem-wide needs. They spoke of working with institutions to provide "*instructional design with AI supports*" or to "*bring best practices to data science education and research.*" Their viewpoint is often from the outside looking in, identifying patterns like the "*siloed kind of implementation*" of AI across campuses or the need for a "*national program for mentorship in ethical AI use.*" They often see their role as a connector, helping to develop "*a more optimized version for your curricula*" by working with both internal and external partners. Their focus was on systemic change, such as building "*a different ecosystem for the infrastructure part*" and funding networks that encouraged collaboration.

## 5. POSITIONALITY CONSIDERATIONS

Experience, discipline, profession, identity, and international background all played a role in experts input for AI education. Below we report on key positionality considerations from our roundtable discussions.

> **Key Ideas**
> 
> 1. **Experience Provides Valuable Context:** Experts who lived through many eras of AI advancement over the previous decades provided valuable historical context and grounded wisdom. New to AI research and early career experts focused more on the immediate challenges of AI education.
> 2. **AI Filtered Through Discipline:** The disciplinary backgrounds of experts often influenced the way they thought about AI education. Experts across computing and humanities fields incorporated areas of their own expertise into their recommendations and priorities around societal impact and data set curation.
> 3. **Professional Role Influences Educational Focus:** Administrative experts (Vice Provost, Associate Director, Dean, Department Chair), tended to think about AI education through the lens of broad, university-level concerns and governance, whereas teaching faculty were more focused on pedagogy, curriculum design, and student experience.
> 4. **Identity Provides Purpose:** Expert experiences with gender, race, culture, and disability provided a grounding for their teaching and goals around AI education, with several citing elements from their own lives as being influential in framing their work for social impact.

## 5.1. Years of Experience

Experts who identified as having worked in AI for decades ("old school AI") sometimes provided historical context, contrasting the current state of generative AI with previous eras of the field. They reminded the group that AI has gone through many cycles and that current trends are built on decades of prior work. This perspective often served to both ground the hype and highlight the truly novel aspects of the current moment. Conversely, experts who were newer to the field or teaching it for the first time often focused on the immediate, practical challenges of designing courses and engaging students in a rapidly changing landscape.

## 5.2. Disciplinary Background

The discussions revealed a clear distinction in framing based on academic discipline. An expert with a background in philosophy consistently emphasized that ethics is a field with its own rigor and expertise, pushing back against the idea that ethical discussions are merely about "vibes." They argued for treating ethics with the same seriousness as technical subjects. An expert from religious studies brought a unique lens to human-robot interaction, analyzing the metaphors and assumptions underlying the technology. Experts from social sciences and education focused on issues of equity, student identity, and culturally responsive pedagogy, with one stating, "*I worry about... opportunity hoarding and access to these kinds of programs.*"

This contrasts with some computing faculty, who tended to frame problems and solutions in more technical or structural terms. For instance, experts with backgrounds in engineering or systems consistently emphasized the importance of implementation, the full software development lifecycle, and understanding how to build robust, scalable, and trustworthy systems. Alternatively, experts with a mathematics or statistics background were often the strongest advocates for a rigorous mathematical foundation to AI-education, believing that a deep understanding of the underlying principles is essential.

Other experts framed AI-education through the particular AI needs of their own disciplines. An instructor from a department focused on Human-Computer Interaction (HCI) emphasized a user-centered and application-focused perspective. Coming from a 20-year industry career in user experience, she expressed skepticism about an AI major that "*feels like it's missing the X, feels like it's missing the concentration or the focus or the application.*" Her background led her to argue that "*computer science is for people,*" and therefore AI education should be grounded in tangible, human-centered problems. In line with this, many HCI scholars highlighted the interdisciplinary nature of their areas and emphasis on ethics. A molecular biologist consistently brought the conversation back to the practical application of AI as a tool within a specific, complex domain. He was less concerned with the nuances of the CS curriculum and more with the end goal: "*what is it you really want them to know?*". He advocated for a focus on critical thinking and the ability to judge the output of AI systems, asking if students can "*tell whether the AI is saying something silly,*" which requires deep domain knowledge, not just technical skill.

## 5.3. Professional Role / Administrative versus Faculty

Experts holding administrative roles (e.g., Vice Provost, Associate Director, Dean, Department Chair) often took a broad, university-wide view to AI-education. Their comments often reflected a high-level, strategic view of the challenges of teaching AI. These experts were often concerned with the logistics of curriculum change, resource allocation, faculty staffing, and the market pressures of competing with other institutions. They were concerned with enrollment numbers, faculty hiring, competition, and institutional politics. One dean explained the rapid development of an AI major as a direct response to a drop in CS freshman enrollment, a decision that "was top driven." A department chair spoke of the administrative burden of creating new programs, noting the extensive reporting and assessment required by the state, which made them reluctant to create new majors or certificates. In contrast, teaching-focused faculty tended to provide specific examples from their own classrooms, detailing assignments, pedagogical techniques, and direct student interactions.

## 5.4. Experience with Underserved Populations/Racial and Gender Identity

Several experts' strong advocacy for fairness was clearly tied to their professional work and personal experiences. A research fellow from an HBCU emphasized the need to understand how AI can be integrated in a way that is "*quite unique to our population*" and the imperative to focus on "student success" for students who may have come from a K-12 school that did not adequately prepare them for college. An expert from an HSI spoke of working with Google on a technology exchange program specifically for such an institution. Another expert, a mother to a nonverbal child who uses an eye-tracking speech device, brought a deeply personal and powerful perspective to the discussion on disability bias, stating:

> "*I see in my own life how problematic it might be for somebody like [my child] to mediate language and communication using [pre-built] AI.*"

This personal connection fueled her call for more research into "neurodiverse data sets" and assistive technologies. A Hawaiian professor — mentioned above — explicitly connected his work to his culture:

> "*I'm building AI systems to help with the revitalization of the Hawaiian language.*"

This positionality reframes the purpose of AI from a generic technological advancement to a tool for cultural preservation and community empowerment. Another expert explicitly centered her contributions on her identity as a Black woman, which directly informed her perspective on the purpose and pedagogy of AI education for her community. She argued that for Black students, the motivation for entering STEM is often social impact, not just technical disruption:

> "*As a black woman on this call, I want my Black students to be able to fine tune, deconstruct models that have been built that are not representative of my community.*"

She further noted that at HBCUs, a majority of first-generation Black students major in the social sciences and humanities, not STEM. This reality, she argued, necessitates a different framing for AI education — one rooted in students interests — to achieve "*immediate adoption from faculty*" and students.

Another expert, leading a "*young women in computing*" program at a Hispanic-serving institution, framed her contributions through a lens focused on building students' feelings of capability in AI. She described her focus as on "*raising that confidence of the students*" to show them that AI is something "*yes, they can*" do, regardless of their major and background. This focus is on developing pedagogical models that allow students to see themselves succeeding in AI work, recognize themselves as potential AI professionals, and feel a sense of belonging in the AI community.

# 6. SUMMARY AND NEXT STEPS

As we work towards building national consensus around AI education, these roundtables have highlighted the need for flexible frameworks that can be adapted to particular institutional contexts. These frameworks would need to offer responses to the questions of financial and computing power resources, pedagogical and curricular resources, and capacity building training for faculty and students. It was highlighted that AI education is not only needed in computing degrees, but in disciplines across industries (e.g., AI + History, AI+ Energy, AI + Health,  AI + Environmental Science), making AI a potential service discipline for the entire university. Additionally — as was brought up by experts in _every_ roundtable — it is important to discuss how computing educators can facilitate discussion and application of ethical AI frameworks and practicums given the broad reaching and integrated impact of these technologies. Overwhelmingly, roundtable experts were excited by the dedicated time to have these conversations and are interested in what cross-institutional partnerships will look like including amongst higher education institutions, research-labs, and industry. One way the Computing Research Association hopes to support these conversations is through our participation in the National Artificial Intelligence Research Resource Pilot Conferences (NAIRR conferences) and National Artificial Intelligence Research Resource Pilot Research Collaboration Network (NAIRR RCN).

We have also compiled and organized AI education resources that were mentioned during our 32 roundtables, linked below. We hope this can contribute to one of the most frequent requests from our roundtable discussions: a central repository of AI education resources for institutions to freely use across higher education.

## 6.1. Next Steps

The insights from the LEVEL UP AI roundtables are informing the LEVEL UP AI workshops, as well as convenings, and resources for the NAIRR Pilot Conferences and NAIRR RCN. Both the Conferences and RCN engage faculty from 1) Research Emerging Institutions Small-Medium, 2) Research Emerging Institutions Large, 3) Historically Black Colleges and Universities, 4) Minority Serving Institutions, 5) Community Colleges, and 6) Four Year Colleges and Universities.

## 6.2. NAIRR Pilot Conferences

The NAIRR Pilot Conferences create opportunities for institutionally-specific convenings around AI in higher education and best practices that are adapted to their particular needs. These conferences will host additional roundtable discussions and in-person workshops in order to build capacity in and access to AI education. Further they create an opportunity to

*Computing Research Association (CRA)*  **36**

build actionable plans for curricular implementation, pedagogical practices, and confidence in educational pathways [13].

## 6.3. NAIRR Pilot RCN

The NAIRR Pilot RCN establishes cross-institutional collaboration and community working towards advancing undergraduate and master's level AI education for all. Faculty who chose to participate in the RCN will share resources, develop AI Education best practices, and establish national strategy through regular convenings, resource development, repositories of best practices, and purposeful engagement towards expanding underrepresented institutions participation in the RCN [12].

## 7. AI EDUCATION RESOURCES

The AI Education Resources collected during the roundtable discussions are available via the shareable link below. The resources on this list should be evaluated by the people using it to determine if a tool or source is useful and in compliance with ethical and regulatory standards at their organization. These do not represent recommendations by CRA in terms of quality, but rather a snapshot of a major landscape of resources documented during the roundtables. Each of the listed resources was grouped into at least one high level category to help with navigating the information. These include: Funding Sources, Open Educational Resources, Educational Resources (paywall), Open Source Technologies, Institutional Infrastructures, Societal Infrastructures, Ethics and Accessibility Resources, Tools, and Books / Articles / Blogs. Institutional Infrastructures are meant to link to things individual universities may be doing in terms of AI major creation curricular pathways or centers for people at those institutions. Societal infrastructures are meant to relate to national and international resources that are accessible to people at a variety of institution types.

Link to **Resources -** **LEVEL-UP AI Resources**

# ACKNOWLEDGEMENTS

*This material is based upon work supported by the U.S. National Science Foundation under Grant No. CNS-2434416. Any opinions, findings, conclusions, or recommendations expressed are those of the authors and do not necessarily reflect the views of the NSF. We thank all of the contributors of these roundtable discussions, which are listed here [insert link]*

In addition to CRA, the following computing organizations support LEVEL UP AI initiative AAAI, ACM, IEEE-CS, and several NSF BPC Alliances.

# CONTRIBUTORS

The authors would like to thank all of the Level UP AI Roundtable participants for sharing their expertise and visions at the intersection of computing and other disciplines to inform this report. A full list of contributors can be found here.